\documentclass[12pt]{article}
\pdfoutput=1
\usepackage[english]{babel}
\usepackage[latin1]{inputenc}
\usepackage{color}
\usepackage[normalem]{ulem}

\usepackage[intlimits]{amsmath}
\usepackage{amssymb}
\usepackage{graphicx}
\usepackage[text={16.4cm,23.7cm}]{geometry}

\begin{document}

\title{\vspace{-2cm} 
{\normalsize
\flushright TUM-HEP 859/12\\}
\vspace{0.6cm} 
\bf Prospects of antideuteron detection \\
from dark matter annihilations or decays\\ 
at AMS-02 and GAPS\\[8mm]}

\author{Alejandro Ibarra and Sebastian Wild\\[2mm]
{\normalsize\it Physik-Department T30d, Technische Universit\"at M\"unchen,}\\[-0.05cm]
{\it\normalsize James-Franck-Stra\ss{}e, 85748 Garching, Germany}
}

\maketitle

\begin{abstract}
The search for cosmic antideuterons has been proposed as a promising method to indirectly detect dark matter, due to the very small background flux from spallations expected at the energies relevant to experiments. The antideuteron flux from dark matter annihilations or decays is, however, severely constrained by the non-observation of an excess in the antiproton-to-proton fraction measured by PAMELA. In this paper we calculate, for representative dark matter annihilation and decay channels, upper limits on the number of antideuteron events at AMS-02 and GAPS from requiring that the associated antiproton flux is in agreement with the PAMELA data. To this end, we first analyze in detail the formation of antideuterons in the coalescence model using an event-by-event Monte Carlo simulation and using data from various high energy experiments. We find that the resulting coalescence momentum shows a dependence on the underlying process and on the center of mass energy involved. Then, we calculate, using a diffusion model, the flux of antideuterons at the Earth from dark matter annihilations or decays. Our results indicate that, despite the various sources of uncertainty, the observation of an antideuteron flux at AMS-02 or GAPS from dark matter annihilations or decays will be challenging.
\end{abstract}

\section{Introduction}

The search for antimatter particles from dark matter annihilations or decays is usually hindered by the existence of large astrophysical backgrounds, which make difficult to unequivocally attribute any possible excess in the fluxes to an exotic origin. A possible exception is the search for cosmic antideuterons. The flux of antideuterons expected from spallations of cosmic rays on the interstellar medium is peaked at a kinetic energy per nucleon $T_{\bar d}\sim 5$ GeV/n and rapidly decreases at smaller kinetic energies~\cite{Duperray:2005si,Donato:2008yx}. In contrast, the spectrum of antideuterons from dark matter annihilations or decays is usually much flatter at low kinetic energies and could easily overcome the astrophysical background~\cite{Donato:1999gy}. For this reason, it has been argued that the search of antideuterons from dark matter annihilations or decays is practically background free and that the observation of one single cosmic antideuteron would constitute an evidence for an exotic antideuteron source. Several authors have calculated the expected antideuteron flux at the Earth in scenarios of dark matter annihilations \cite{Donato:1999gy,Baer:2005tw,Brauninger:2009pe} or decays \cite{Ibarra:2009tn}, using a factorized approximation of the coalescence model; these analyses have been recently revisited in \cite{Kadastik:2009ts,Cui:2010ud,Dal:2012my} using an event-by-event simulation.

Antideuterons are produced by the coalescence of one antiproton and one antineutron. It is then apparent that there is a strong correlation between the cosmic antideuteron flux and the cosmic antiproton flux. In this paper we will analyze the prospects to detect primary antideuterons at AMS-02 \cite{Ahlen:1994ct} or GAPS \cite{Mori:2001dv}, in view of the existing limits on the dark matter annihilation cross section or decay width into antiprotons from the PAMELA data on the cosmic antiproton-to-proton fraction~\cite{Adriani:2010rc}. We will show that, although the limits on the cross section or the decay width that follow from the non-observation on an excess in the measured antiproton-to-proton fraction are very sensitive to the choice of the propagation parameters and, to a lesser extent, to the choice of the dark matter halo profile, due to the correlation between the antideuteron and the antiproton production and propagation mechanisms, most of the uncertainties cancel out when calculating an upper limit on the antideuteron flux and the number of antideuteron events, thus giving fairly robust conclusions.

We will analyze in section \ref{sec:coalescence} the antideuteron production employing the coalescence model, which has only one free parameter, the coalescence momentum. We will determine this parameter using an event-by-event Monte Carlo simulation, analyzing data from various experiments measuring antideuteron or deuteron production. Our calculation suggests that the coalescence momentum is not universal, but depends on the underlying process and on the center of mass energy. Using the same event-by-event Monte Carlo, we then calculate in section \ref{sec:DM} an upper limit on the antideuteron flux and the number of events at AMS-02 and GAPS for two representative scenarios of dark matter annihilation and decay, imposing the requirement that the cosmic antiproton-to-proton ratio is in agreement with the PAMELA measurements. Lastly, in section \ref{sec:conclusions} we will present our conclusions.

\section{The coalescence model}
\label{sec:coalescence}
To describe the antideuteron production we will employ the coalescence model~\cite{Butler:1963pp,Schwarzschild:1963zz,Csernai:1986qf,Chardonnet:1997dv,Kadastik:2009ts}, which postulates that the probability of the formation of an antideuteron out of an antiproton-antineutron pair with given four-momenta $k_{\bar{p}}^{\mu}$ and $k_{\bar{n}}^{\mu}$ can be approximated as a narrow step function $\Theta \left( \Delta^2+p_0^2 \right)$, where $\Delta^{\mu} = k_{\bar{p}}^{\mu} - k_{\bar{n}}^{\mu}$. In this model, the coalescence momentum $p_0$ is the maximal relative momentum of the two antinucleons that still allows the formation of an antideuteron.

One can show~\cite{Kadastik:2009ts} that for $| \vec{k}_{\bar d}| \gg p_0$, being $ \vec{k}_{\bar d}=\vec k_{\bar{p}}+ \vec k_{\bar{n}}$, this ansatz leads to the following differential antideuteron yield in momentum space:
\begin{align}
\label{eqn:general_coal_formula}
 \gamma_{\bar d} \, & \frac{d^3N_{\bar d}}{d^3 k_{\bar d}} ( \vec{k}_{\bar d}) = \frac18 \cdot \frac43 \pi p_0^3 \cdot  \gamma_{\bar{p}} \gamma_{\bar{n}} \frac{d^3 N_{\bar{p}} d^3 N_{\bar{n}}}{d^3 k_{\bar{p}} d^3 k_{\bar{n}}} \left( \frac{\vec{k}_{\bar d}}{2}, \frac{\vec{k}_{\bar d}}{2} \right) \, ,
\end{align}
so the number of antideuterons scales with $p_0^3$. Note furthermore, that under the assumption of statistical independence of the antiproton and antineutron production, Eq. (\ref{eqn:general_coal_formula}) reduces again to the formula used in the factorized approximation of the coalescence model~\cite{Donato:1999gy,Baer:2005tw,Brauninger:2009pe,Ibarra:2009tn}.\footnote{In some works the factor $1/8$ appearing in Eq. (\ref{eqn:general_coal_formula}) is included in the definition of the coalescence momentum, resulting in a value of $p_0$ which is one half of the coalescence momentum defined through Eq. (\ref{eqn:general_coal_formula}).} In absence of a microscopic understanding of the coalescence mechanism, the coalescence momentum $p_0$ should be determined from experiments.

We have analyzed various high-energy processes of antideuteron production and we have implemented the coalescence model in an event-by-event simulation. Concretely, we have used the Monte Carlo generators PYTHIA 6~\cite{Sjostrand:2006za} or PYTHIA 8~\cite{Sjostrand:2007gs} to simulate the processes and we have selected those events with present an antiproton-antineutron pair in the final state with a relativistic invariant momentum difference $-\Delta^2 < p_0^2$. Furthermore, we have also required that the antiproton and the antineutron are close in position space, as the nuclear forces that are responsible for the binding of the antideuteron have a finite range in the order of a few fm. Hence, we have included in our analysis the antiprotons and antineutrons produced directly in the hadronization process as well as those produced in strong decays, such as from $\bar \Delta$ decay which promptly decay with a lifetime $\tau_\Delta=5\times 10^{-24}$ s, very close to the primary vertex. In contrast, we have neglected those produced in weak decays, such as from $\bar \Lambda$ and $\bar \Sigma^\pm$ decay, which have a lifetime $\tau_{\Lambda,\Sigma}\sim 10^{-10}$ s and which then have a macroscopic decay length, much larger than the typical range of the nuclear forces. In consequence, antinucleons produced in weak decays have no chance to coalesce with antinucleons produced in the hadronization or in other weak decays, even if their relative momentum is smaller than $p_0$.

\begin{figure}
\begin{center}
\includegraphics[bb=0 0 330 240,width=0.45\textwidth]{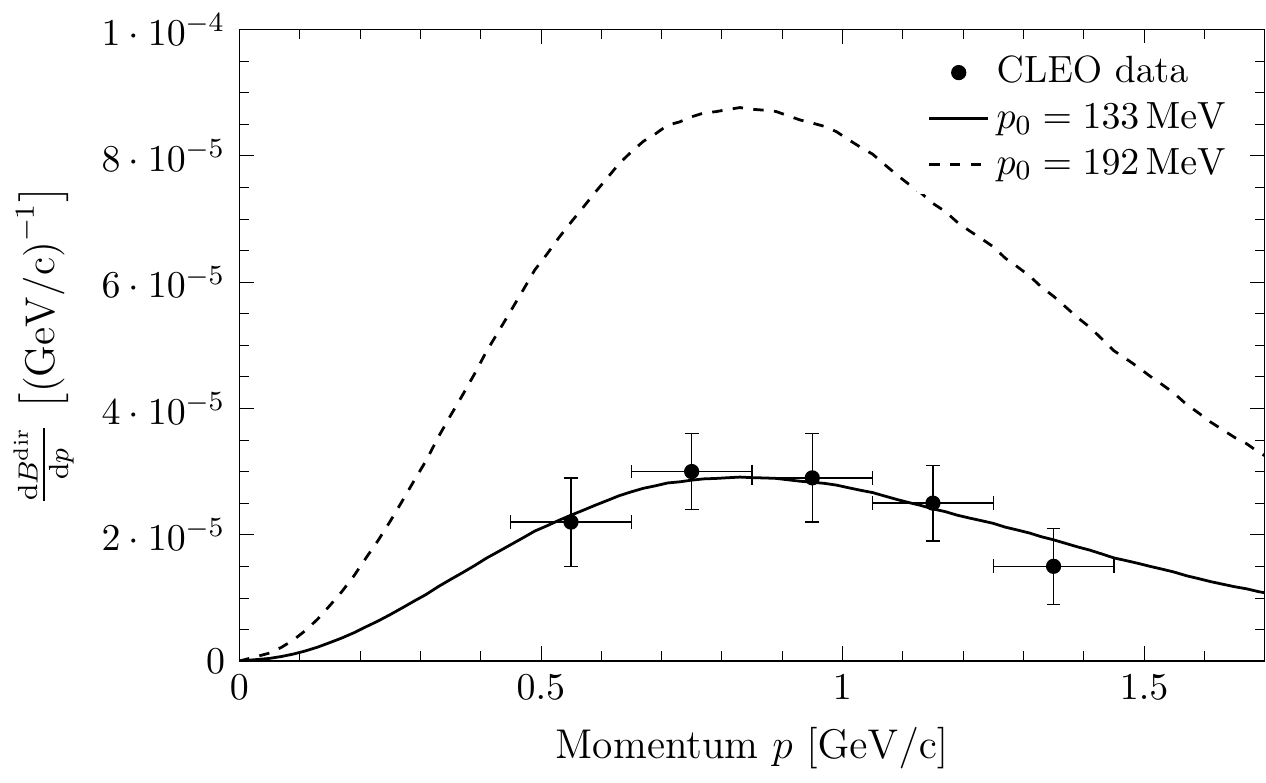}\hspace{.8cm}
\includegraphics[bb=0 0 360 240,width=0.49\textwidth]{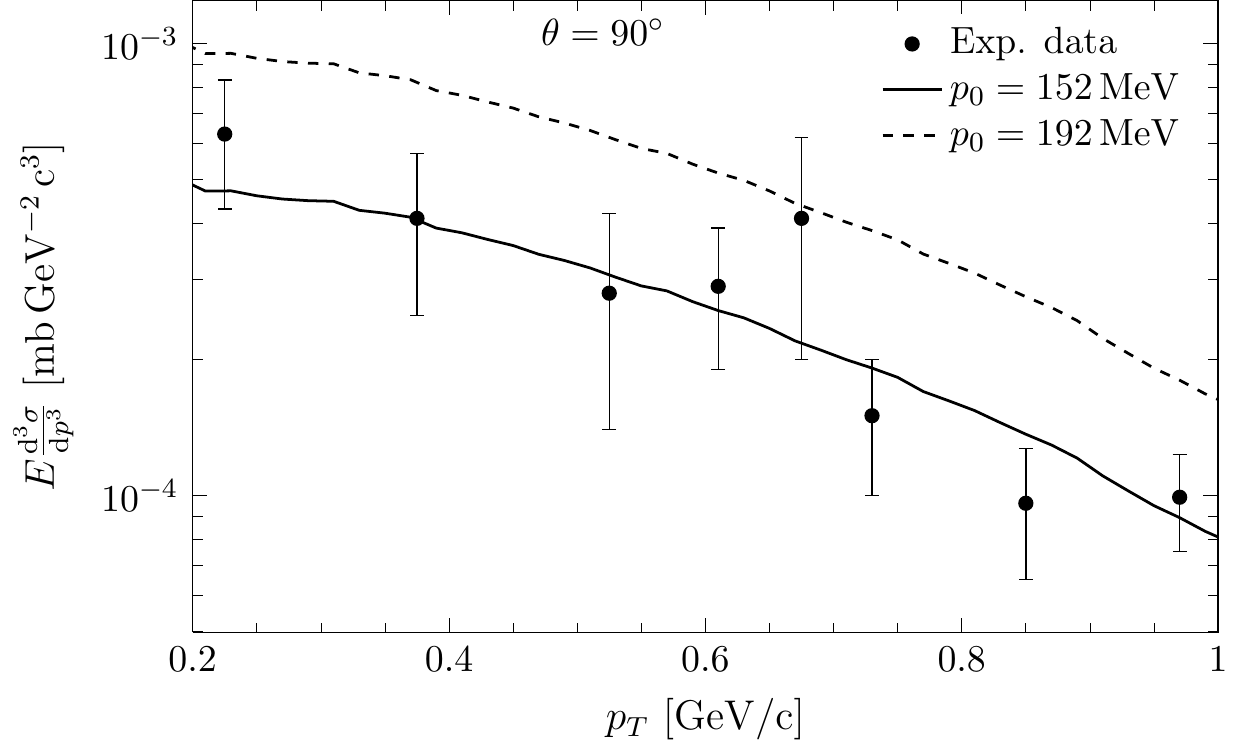}\\\hspace{-.4cm}
\includegraphics[bb=00 0 340 240,width=0.46\textwidth]{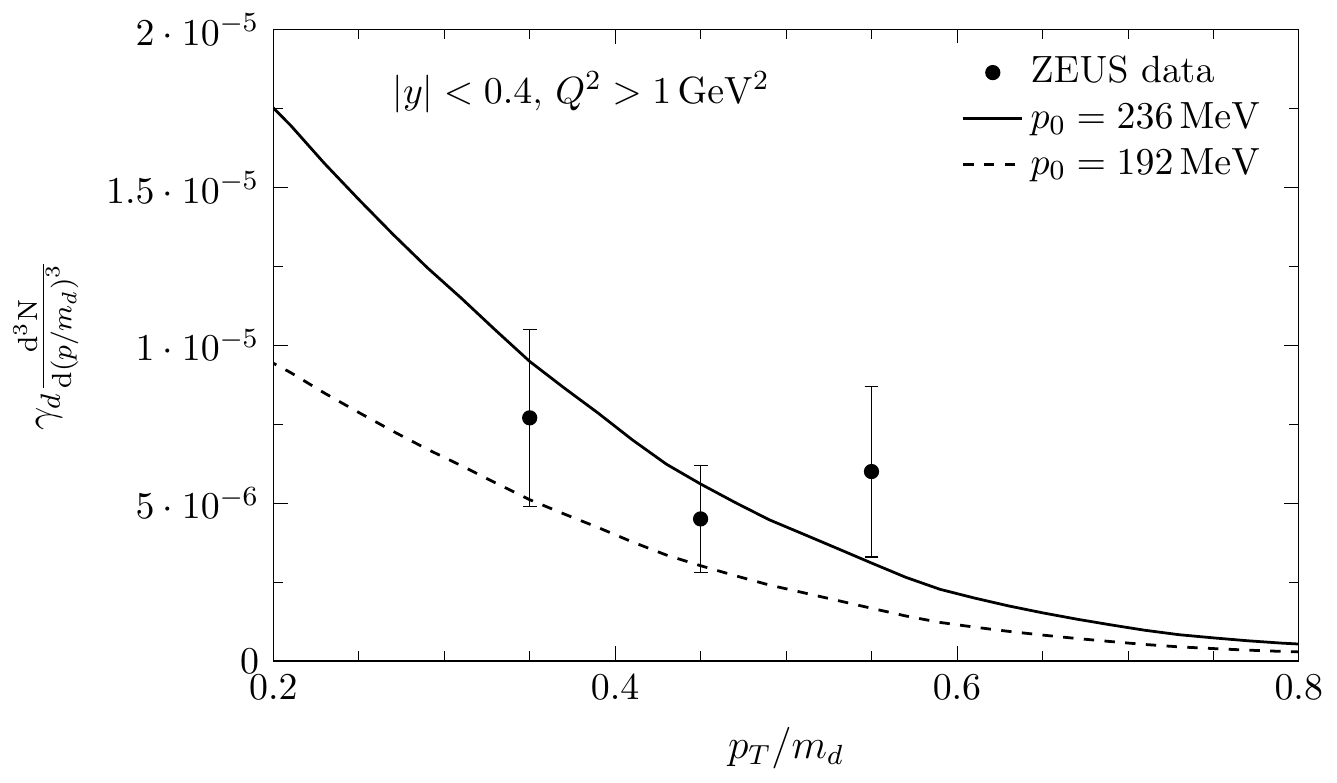}\hspace{.7cm}
\includegraphics[bb=0 0 360 240,width=0.49\textwidth]{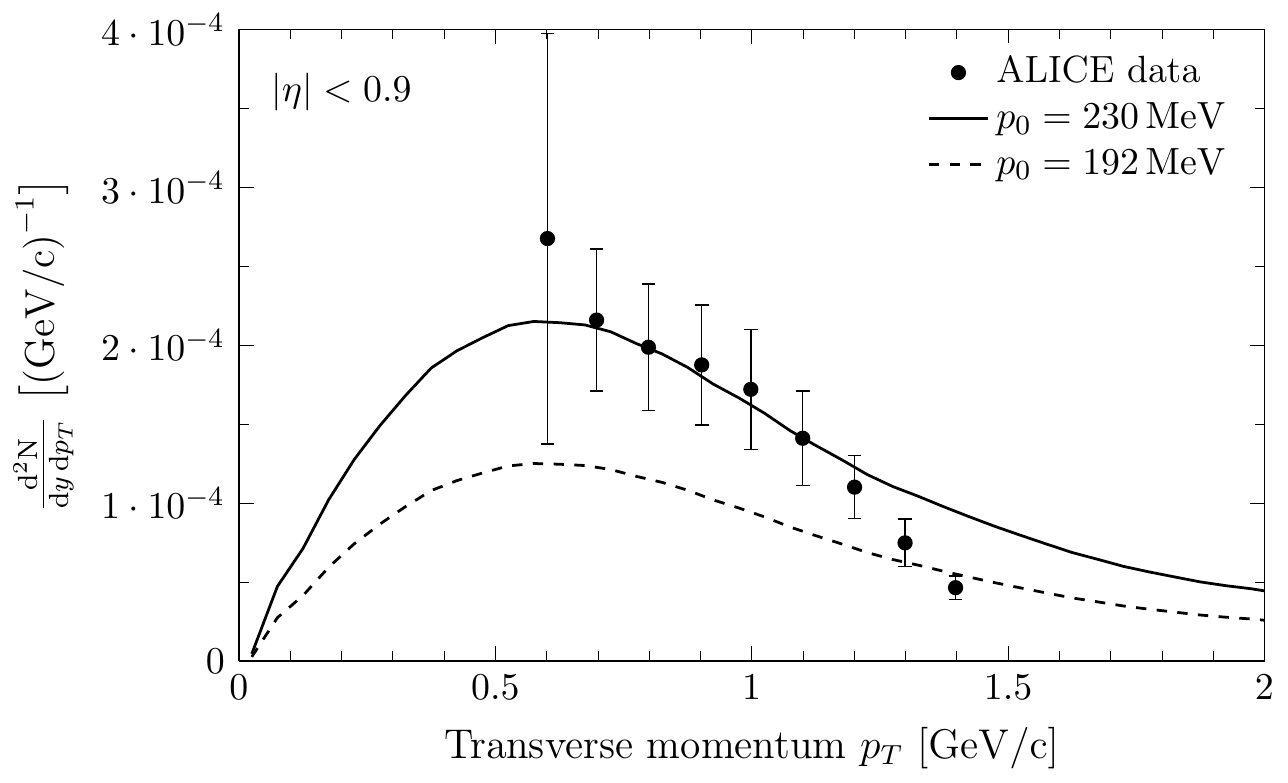}
\caption{Antideuteron/deuteron spectra measured in various experiments compared to the prediction of the coalescence model for a value of the coalescence momentum $p_0=192$ MeV (inferred from the ALEPH measurement of the antideuteron multiplicity in $Z$ boson decay) and for the best fit value for each experiment. {\it Top left}: antideuteron spectrum measured at CLEO from $\Upsilon$ decay, {\it top right}: antideuteron spectrum measured at CERN ISR from $pp$ collisions at $\sqrt{s}=53$ GeV, {\it bottom left}: antideuteron spectrum measured at ZEUS from $e^{-} p$ collisions at $\sqrt{s}=318$ GeV, {\it bottom right}: preliminary deuteron spectrum measured by ALICE in $pp$ collisions at $\sqrt{s}=7$ TeV. See the text for details.}
\label{fig:coalescence}
\end{center}
\end{figure}

The processes we have analyzed and the corresponding coalescence momentum inferred from the procedure described above are:
\begin{itemize}
\item Multiplicity of antideuterons from $Z$ boson decay measured by ALEPH \cite{Schael:2006fd}. The number of antideuterons per hadronic $Z$ decay with a momentum between 0.62 and 1.03 GeV/c and a production angle satisfying $\left| \cos \theta \right| < 0.95$ was measured to be $\left( 5.9 \pm 1.8 \pm 0.5 \right) \times 10^{-6}$. We reproduce this number by choosing $p_0=$ 192 MeV in our simulation (using PYTHIA 8); the quoted error can be translated to a uncertainty of $\sim$ 30 MeV in the coalescence momentum.\footnote{In \cite{Kadastik:2009ts,Cui:2010ud} a value of $p_0=$ 160 MeV is extracted from the $Z$ decay measurement. However, this value is obtained by allowing also the antiprotons and antineutrons originating from weak decays to coalesce.}

\item Antideuteron production in the decay of the $\Upsilon \left( \mbox{1S} \right)$ meson produced in $e^{+} \, e^{-}$ collisions, measured by CLEO \cite{Asner:2006pw}. In the data and therefore also in our simulation (using PYTHIA 8), the number of antideuterons is normalized to the number of $\Upsilon \left( \mbox{1S} \right)$ decaying to either $g \, g \, g$ or $g \, g \, \gamma$, denoted as 'direct' decays in \cite{Asner:2006pw}. In Fig. \ref{fig:coalescence}, top left plot, we show the data and the results of our simulation for the direct branching ratio $\frac{d B^{\text{dir}}}{d p}$ into antideuterons, being the best-fit value of the coalescence momentum $p_0=$133 MeV.

\item Antideuteron production in $pp$ collisions at $\sqrt{s}=53$ GeV at the CERN ISR \cite{Alper:1973my,Henning:1977mt}. Here the invariant differential cross section for the inclusive $\bar d$ production at $\theta_{\text{CM}}=90^{\circ}$ was measured as a function of the transverse momentum. As shown in Fig. \ref{fig:coalescence}, top right plot, we reproduce the spectrum with PYTHIA 8 by choosing $p_0=$152 MeV.

\item Antideuteron production in $e^-p$ collisions at $\sqrt{s}=318$ GeV measured by the ZEUS collaboration \cite{Chekanov:2007mv}. The experimental analysis as well as our simulation was performed for deep inelastic scattering events with an exchanged photon virtuality $Q^2 > 1$ GeV$^2$. The measured quantity is the Lorentz invariant differential antideuteron yield in the rapidity range $\left| y \right| < 0.4$. The three data points are shown in Fig. \ref{fig:coalescence}, bottom left plot, together with our simulation using a best-fit value of $p_0=236$ MeV. Here we used PYTHIA 6, as deep inelastic scattering is not included in PYTHIA 8 yet.

\item Preliminary data on deuteron production in $pp$ collisions at $\sqrt{s}=7$ TeV by ALICE \cite{Sharma:2012zz}. At very high energies, the production cross sections for protons and antiprotons are nearly equal (see e.g. \cite{:2012qb}); it is therefore reasonable to assume the same for the production cross sections of deuterons and antideuterons which allows us to extract information about the coalescence momentum from this measurement of deuterons. The double differential deuteron yield $\frac{d^2 N}{dy \, dp_T}$ was measured as a function of the transverse momentum in the pseudorapidity range $\left| \eta \right| < 0.9$ and is shown in Fig. \ref{fig:coalescence}, bottom right plot, together with our simulation using the best-fit value of $p_0=230$ MeV. As discussed in \cite{:2012qb}, PYTHIA 6 (tune D6T) reproduces the (anti)proton spectrum better than PYTHIA 8, therefore we used the former for our simulation.
\end{itemize}

We have checked, for each data set, that the Monte Carlo generator used for the determination of the coalescence momentum reproduces reasonably well the corresponding antiproton spectrum.

As apparent from the plots, where we also show the predicted spectra for the value of the coalescence momentum inferred from $Z$ decays, $p_0=192$ MeV, it is impossible to reproduce the results of all these experiments with a single coalescence momentum. Understanding the dependence of the coalescence momentum with the underlying hard process and with the center of mass energy is in our opinion an interesting problem which deserves further investigation, nevertheless in this paper will not pursue this direction and instead we will concentrate on the indirect detection of antideuterons from dark matter annihilations or decays, bearing in mind the above-mentioned limitation of the coalescence model to describe the production of antideuterons. 

\begin{figure}
\begin{center}
\includegraphics[bb=0 0 360 240,width=0.48\textwidth]{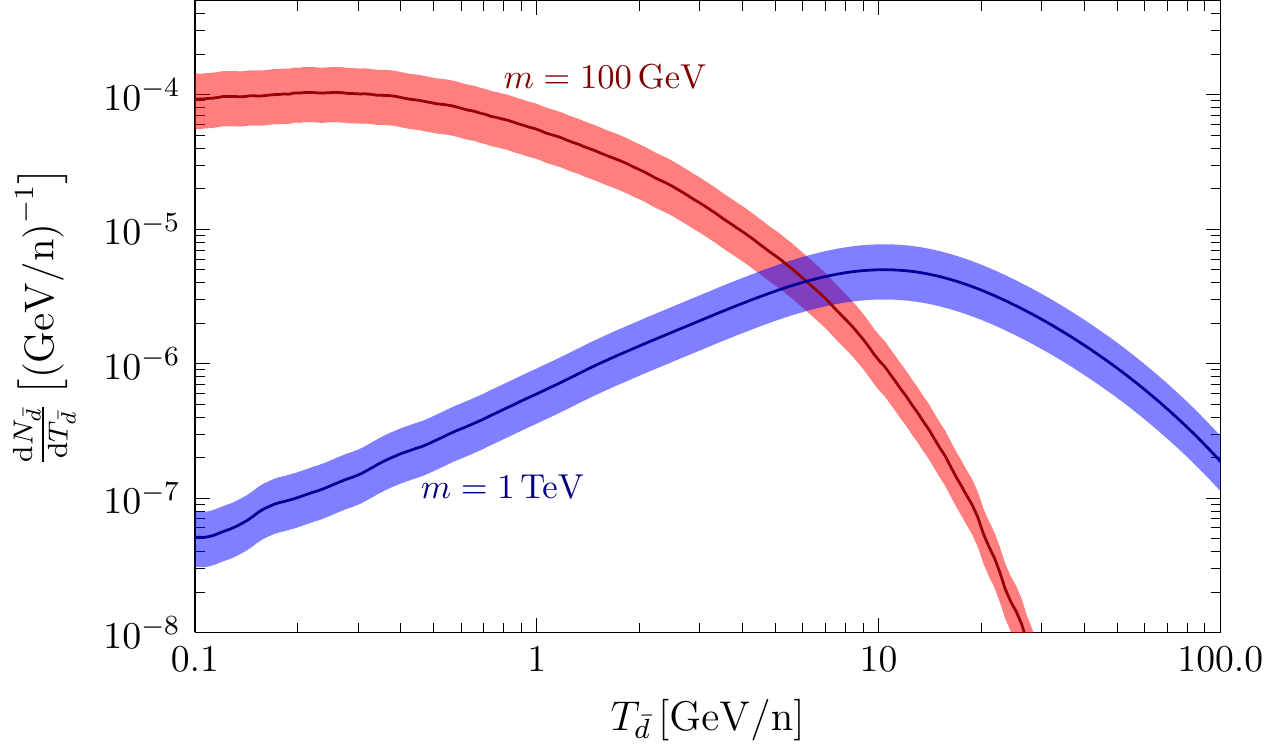}
\includegraphics[bb=0 0 360 240,width=0.48\textwidth]{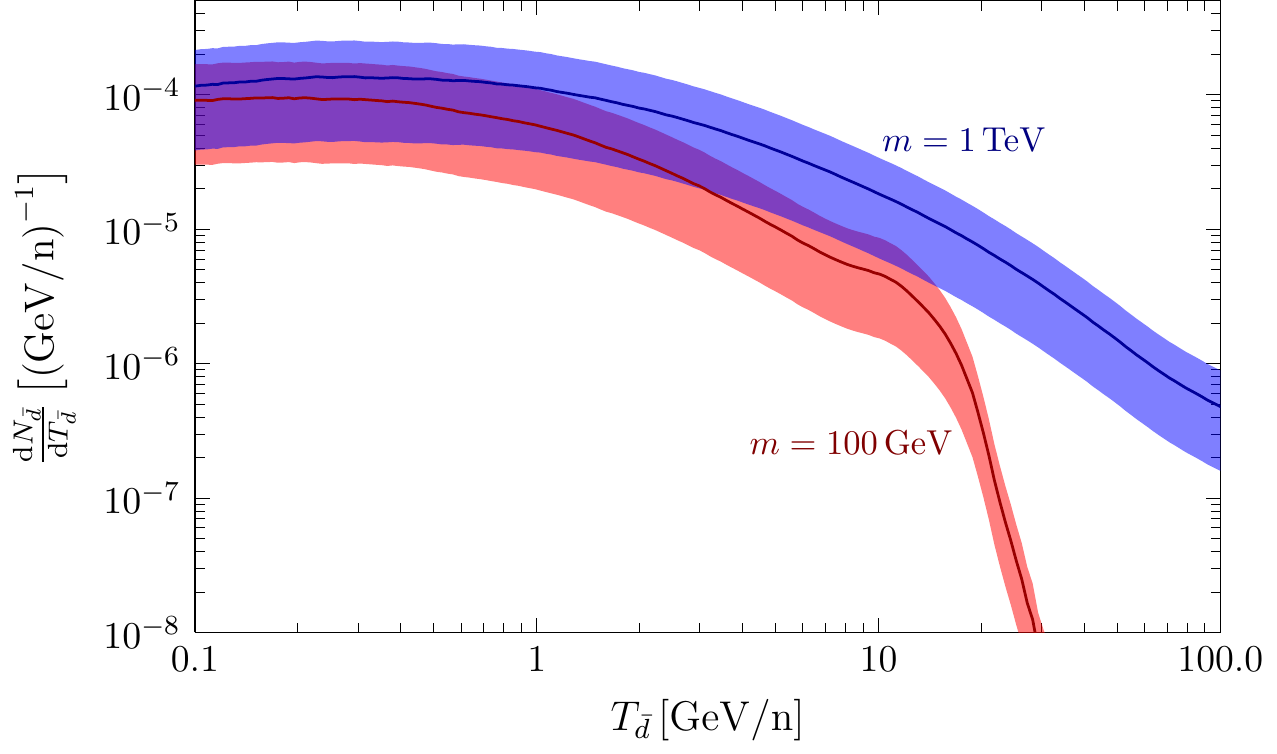}
\caption{Antideuteron injection spectra $\frac{d N_{\bar d}}{d T_{\bar d}}$ expected from dark matter annihilations into $W^+ W^-$ (left plot) and $b \bar b$ (right plot) for dark matter masses $m_{\rm DM}=100$ GeV and $m_{\rm DM}=1$ TeV. The solid line corresponds to a value of the coalescence momentum $p_0=192$ MeV, while the shaded regions show the uncertainty band corresponding to the ranges  $162\,{\rm MeV} \leq p_0 \leq 222\,{\rm MeV}$ and $133\,{\rm MeV} \leq p_0 \leq 236\,{\rm MeV}$ for annihilations into $W^+ W^-$ and $b\bar b$ respectively.}
\label{fig:Spectra}
\end{center}
\end{figure}

We will analyze in this paper the antideuteron flux from dark matter annihilations or decays into $W^+ W^-$ or into $b\bar b$. In both cases, we will take the value of the coalescence momentum inferred from $Z$ decays, $p_0=192$ MeV. This choice is well justified for dark matter annihilations/decays into $W^+ W^-$, irrespectively of the dark matter mass. However, as argued above, the coalescence momentum for dark matter annihilations or decays directly into partons can differ from this value. To estimate the impact of the uncertainties in the coalescence model on the cosmic antideuteron fluxes, we show in Fig. \ref{fig:Spectra}, left plot, the range of antideuteron injection spectra from ${\rm DM}\,{\rm DM}\rightarrow W^+ W^-$ for dark matter masses $m_{\rm DM}=100$ GeV and $m_{\rm DM}=1$ TeV taking $p_0$ between 162 MeV and 222 MeV, the range that results from the uncertainty in the determination of the coalescence momentum from the ALEPH data. We also show, in the right plot, the uncertainty in the antideuteron spectrum for  ${\rm DM}\,{\rm DM}\rightarrow b\bar b$, also for $m_{\rm DM}=100$ GeV and $m_{\rm DM}=1$ TeV, taking $p_0$ between 133 MeV and 236 MeV, which is the range of coalescence momenta we find from our analysis of the experimental data on antideuteron and deuteron production.

\section{Antideuterons from dark matter annihilations/decays}
\label{sec:DM}

We will assume that the Milky Way is embedded in a spherically symmetric dark matter halo with radial distribution $\rho(r)$. In order to evaluate the uncertainty in the predictions of the antideuteron flux from our ignorance of the structure of the dark matter halo, we will show results for the Navarro-Frenk-White (NFW) density profile~\cite{Navarro:1995iw,Navarro:1996gj}:
\begin{equation}
  \rho_\text{DM}(r)=\frac{\rho_0}{(r/r_\text{s})
  [1+(r/r_\text{s})]^2}\;,
\end{equation}
with scale radius $r_s = 24.42 ~\rm{kpc}$, the Einasto profile~\cite{Navarro:2003ew,Graham:2005xx,Navarro:2008kc}:
\begin{equation}
  \rho_\text{DM}(r)=\rho_0 \exp\left\{-\frac{2}{\alpha}
\left[\left(\frac{r}{r_s}\right)^\alpha-1\right]\right\}\;,
\end{equation}
with $\alpha=0.17$ and $r_s=28.44 ~\rm{kpc}$, and the much shallower isothermal profile~\cite{Bahcall:1980fb}:
\begin{equation}
  \rho_\text{DM}(r)=\frac{\rho_0 }{r^2+r_s^2}\;,
\end{equation}
with $r_s=4.38~\rm{kpc}$. In all the cases, the overall normalization factor $\rho_0$ will be chosen to reproduce the local dark matter density $\rho_\odot= 0.39~\text{GeV}/\text{cm}^3$~\cite{Catena:2009mf,Weber:2009pt,Salucci:2010qr,Pato:2010yq,Iocco:2011jz} with $r_\odot=8.5$ kpc.

The annihilation or decay of dark matter particles with mass $m_{\rm DM}$ at the position $\vec{r}$ with respect to the center of the Galaxy produces antideuterons with a rate per unit of kinetic energy and unit volume given by
\begin{align}
Q_{\bar d}(T,\vec r)=
\begin{cases} \displaystyle{ \frac{1}{2}\frac{\rho_{\rm DM}^2(\vec r)}{m^2_{\rm DM}}
\sum_f \langle \sigma v\rangle_f \frac{dN^f_{\bar d}}{dT}} & {\rm (annihilations)}\;, \\
\displaystyle{\frac{\rho_\text{DM}(\vec{r})}{m_{\rm DM}}\sum_f \Gamma_f\frac{dN^f_{\bar d}}{dT}} & {\rm (decays)}\;,
\end{cases}
\label{eq:source}
\end{align}
where $\langle \sigma v\rangle_f$ denotes the velocity weighted annihilation cross-section, $\Gamma_f$ the decay rate and $dN^f_{\bar d}/dT$ the energy spectrum of antideuterons produced in the annihilation or decay channel $f$. The antideuteron injection spectra $dN_{\bar d}/dT$ are calculated with PYTHIA 8 using the event-by-event coalescence model with $p_0=$ 192 MeV, as discussed in the previous chapter.

After being produced, antideuterons propagate in a complicated way before reaching the Earth. We will simulate, following \cite{ACR}, the antideuteron propagation by means of a stationary two-zone diffusion model with cylindrical boundary conditions. Then, the number density of antideuterons per unit kinetic energy at the position $\vec r$, $f_{\bar d}(T,\vec{r})$, satisfies the following transport equation:
\begin{equation}
0=\frac{\partial f_{\bar d}}{\partial t}=
\nabla \cdot (K(T,\vec{r})\nabla f_{\bar d})
-\nabla \cdot (\vec{V_c}(\vec{r})  f_{\bar d})
-2 h \delta(z) \left( \Gamma_{\rm ann} + \Gamma_{\rm non-ann} \right) f_{\bar d}+Q_{\bar d}(T,\vec{r})\;.
\label{transport-antip}
\end{equation}
In this equation energy losses and reacceleration have been neglected. We will impose as boundary conditions that the  number density of antideuterons vanishes at the boundary of the diffusion zone, which we approximate by a cylinder with half-height $L = 1-15~\rm{kpc}$ and radius $ R = 20 ~\rm{kpc}$.

The first term on the right-hand side of the transport equation is the diffusion term, which accounts for the propagation through the tangled Galactic magnetic field. The diffusion coefficient $K(T,\vec{r})$ is assumed to be constant throughout the diffusion zone and is parametrized by:
\begin{equation}
K(T)=K_0 \;\beta\; {\cal R}^\delta\;,
\end{equation}
where $\beta=v/c$ and ${\cal R}$ is the rigidity of the particle, which is defined as the momentum in GeV per unit charge, ${\cal R}\equiv p({\rm GeV})/Z$. The normalization $K_0$ and the spectral index $\delta$ of the diffusion coefficient are related to the properties of the interstellar medium and can be determined from the flux measurements of other cosmic ray species, mainly from the Boron to Carbon (B/C) ratio~\cite{Maurin:2001sj}. The second term is the convection term, which accounts for the drift of charged particles away from the disk, induced by the Milky Way's Galactic wind. It has axial direction and is also assumed to be constant inside the diffusion region: $\vec{V}_c(\vec{r})=V_c\; {\rm sign}(z)\; \vec{k}$. The ranges of the astrophysical parameters $\delta$, $K_0$, $L$ and $V_c$ that are consistent with the B/C ratio and that produce the minimal (MIN), median (MED) and maximal (MAX) antideuteron fluxes  were derived in \cite{Donato:2003xg} and are listed in Table \ref{tab:param-antideuteron}. Lastly, the third part describes annihilations of antideuterons with ordinary matter in the Galactic disk, which is assumed to be an ``infinitely'' thin disk with half-width $h=100$ pc, and  non-annihilating interactions of antideuterons with interstellar matter. In the latter processes antideuterons lose energy, leading to a redistribution of the flux towards lower energies. However, this ``tertiary'' contribution to the low energy part of the spectrum turns out to be negligibly small for primary antideuterons, hence we will treat these antideuterons as if they simply disappear from the flux.
The reaction rates for these two processes can be written as
\begin{equation}
\Gamma_{\rm ann} + \Gamma_{\rm non-ann}=(n_{\rm H}+4^{2/3} n_{\rm He})
\left( \sigma^{\rm ann}_{\bar d p} + \sigma^{\rm non-ann}_{\bar d p} \right) v \;,
\end{equation}
where $n_{\rm H}\sim 1\,{\rm cm}^{-3}$ is the number density of Hydrogen nuclei in the Milky Way disk and $n_{\rm He}\sim 0.07 \,n_{\rm H}$ the number density of Helium nuclei. In this expression it has been assumed that the inelastic cross section between an antideuteron and a helium nucleus is related to the inelastic cross section between an antideuteron and a proton by the simple geometrical factor $4^{2/3}$. Following~\cite{Donato:2008yx}, we use for our numerical analysis the estimation  $\sigma^{\rm inel}_{{\bar d} p} \left( T_{\bar d} / n \right) \simeq 2\sigma^{\rm inel}_{{\bar p}p} \left( T_{\bar{p}} = T_{\bar d} /  n \right)$, where $\sigma^{\rm inel} = \sigma^{\rm ann} + \sigma^{\rm non-ann}$ and $T_{\bar d} / n$ is the kinetic energy per nucleon of the antideuteron. The inelastic antiproton-proton cross section is taken from \cite{Tan:1983de}:
\begin{equation}
\sigma^{\rm inel}_{{\bar p}p} \left( T_{\bar{p}} \right) = 24.7 \, \left( 1 + 0.584 \, T_{\bar{p}}^{-0.115} + 0.856 \, T_{\bar{p}}^{-0.566} \right) \, {\rm mbarn}\;,
\end{equation} 
with $T_{\bar{p}}$ in units of GeV. 

\begin{table}[t]
\begin{center}
\begin{tabular}{|c|cccc|}
\hline
Model & $\delta$ & $K_0\,({\rm kpc}^2/{\rm Myr})$ & $L\,({\rm kpc})$
& $V_c\,({\rm km}/{\rm s})$ \\
\hline 
MIN & 0.85 & 0.0016 & 1 & 13.5 \\
MED & 0.70 & 0.0112 & 4 & 12 \\
MAX & 0.46 & 0.0765 & 15 & 5 \\
\hline
\end{tabular}
\caption{\label{tab:param-antideuteron}  Astrophysical parameters compatible with the B/C ratio that yield the minimal (MIN), median (MED) and maximal (MAX) antideuteron flux~\cite{Donato:2003xg}.}
\end{center}
\end{table}

Finally, the interstellar flux of primary antideuterons at the Solar System from dark matter annihilation or decays is given by:
\begin{equation}
\Phi^{\rm{IS}}_{\bar d}(T) = \frac{v}{4 \pi} f_{\bar d}(T,r_\odot),
\label{flux}
\end{equation}
where $v$ is the antideuteron velocity. This is not, however, the antideuteron flux measured by experiments, which is affected by solar modulation. In the force field approximation~\cite{solar-modulation} the effect of solar modulation can be included by applying the following simple formula that relates the antideuteron flux at the top of the Earth's atmosphere and the interstellar antideuteron flux~\cite{perko}:
\begin{equation}
\Phi_{\bar d}^{\rm TOA}(T_{\rm TOA})=
\left(
\frac{2 m_D T_{\rm TOA}+T_{\rm TOA}^2}{2 m_D T_{\rm IS}+T_{\rm IS}^2}
\right)
\Phi_{\bar d}^{\rm IS}(T_{\rm IS}),
\label{eq:solar-modulation}
\end{equation}
where $T_{\rm IS}=T_{\rm TOA}+\phi_F$, with $T_{\rm IS}$ and $T_{\rm TOA}$ being the antideuteron kinetic energies at the heliospheric boundary and at the top of the Earth's atmosphere, respectively, and $\phi_F$ being the Fisk potential, which varies between 500 MV and 1.3 GV over the eleven-year solar cycle. For our numerical analysis we will choose $\phi_F=500$ MV.

A flux of antideuterons from dark matter annihilations or decays is necessarily accompanied by a flux of antiprotons. Hence, the non-observation by the PAMELA collaboration of an excess in the antiproton-to-proton fraction over the expected astrophysical background can be used to set upper limits on the antideuteron flux from dark matter annihilations or decays. To this end, we first calculate, for a given dark matter annihilation or decay channel, an upper limit on the dark matter annihilation cross section or decay width using the same formalism as for antideuterons, with the obvious changes in the particle mass and $\sigma_{\bar d p}^{\text{inel}} \rightarrow \sigma_{\bar{p} p}^{\text{inel}}$ (see also \cite{Cholis:2010xb,Garny:2012vt,Tavakoli:2011wz}). Then, we translate the upper limit on the annihilation cross section in that channel into an upper limit on the antideuteron flux, calculated following the procedure explained above. Upper limits on the antideuteron flux from decaying dark matter are obtained in an analogous way.

\begin{figure}
\begin{center}
\includegraphics[bb=0 0 360 240,width=0.48\textwidth]{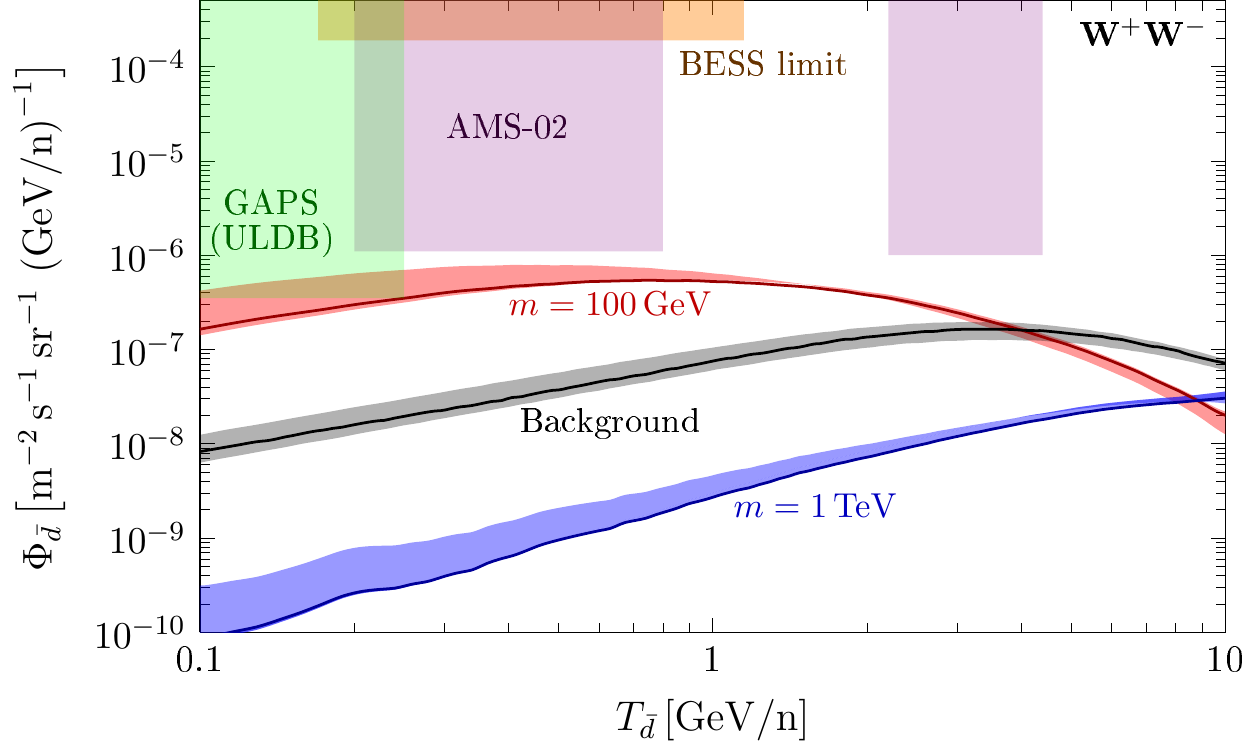}
\includegraphics[bb=0 0 360 240,width=0.48\textwidth]{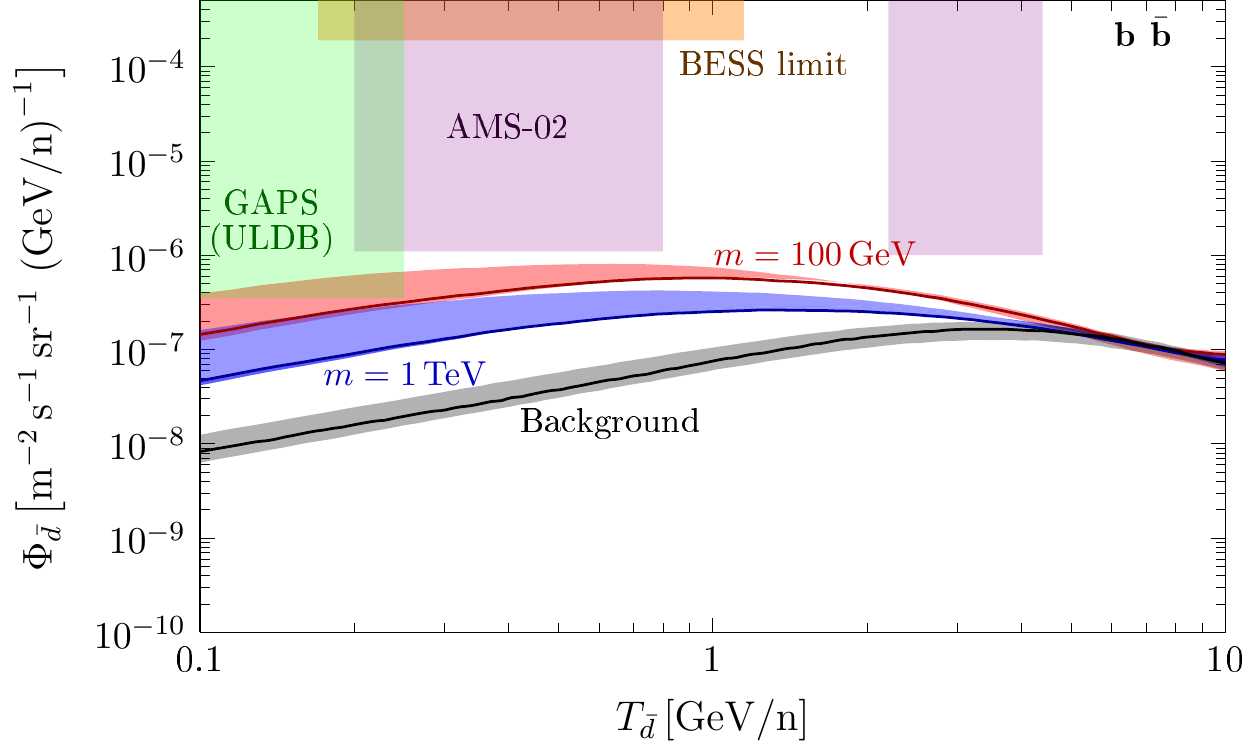}
\caption{Maximum antideuteron flux from dark matter annihilations into $W^+W^-$ (left plot) and $b \bar b$ (right plot) compatible with the PAMELA measurements of the antiproton-to-proton fraction for dark matter masses $m_{\rm DM}=100$ GeV (red line) and $m_{\rm DM}=1$ TeV (blue line) assuming a NFW dark matter halo profile, the MED propagation model and a value of the coalescence momentum $p_0=192$ MeV. We also show as a black line the expected background flux calculated in~\cite{Donato:2008yx}, also for the MED propagation model. The red, blue and grey shaded regions span the propagation uncertainty between MIN, MED and MAX parameters.}
\label{fig:fluxes}
\end{center}
\end{figure}

 We show in Fig. \ref{fig:fluxes} the maximal antideuteron fluxes from dark matter annihilations into $W^+W^-$ and into $b\bar b$ which are consistent with the PAMELA measurements of the antiproton-to-proton fraction for $m_{\rm DM}=100$ GeV and 1 TeV, as well as the predicted cosmic antideuteron background from spallation of cosmic rays on the interstellar medium; for the latter, we use the calculation presented in~\cite{Donato:2008yx}. The band reflects the uncertainty in the prediction due to the choice of the propagation model.\footnote{\label{footnote-had}The background flux also suffers from nuclear uncertainties which conservatively amount to one order of magnitude below 5 GeV, which is the energy range relevant for AMS-02 and GAPS~\cite{Donato:2008yx}. We remark that the background uncertainties do not modify the primary antideuteron fluxes but only affect the significance of a possible dark matter signal.} It is important to note that, despite the predictions for the antiproton and the antideuteron fluxes at Earth can vary by almost two orders of magnitude depending on the propagation model, when normalizing the primary antideuteron flux to the primary antiproton flux a large part of the uncertainty cancels out, due to the fact that the propagation of antideuterons is tightly correlated to the propagation of antiprotons. As a result, the uncertainty in the maximal antideuteron flux allowed by the measured $\bar p/p$ ratio amounts to less than a factor of four. It is also apparent from the plot that there is a strong dependence of the maximal antideuteron flux from ${\rm DM}\,{\rm DM} \rightarrow W^+W^-$ with the dark matter mass, in contrast to that from ${\rm DM}\,{\rm DM}\rightarrow b\bar b$, due to different shape of the corresponding antideuteron injection spectra ({\it c.f.} Fig.~\ref{fig:Spectra}). 

We also show in the plot the best present limit on the cosmic antideuteron flux, which was set by BESS in the range of kinetic energy per nucleon $0.17\leq T\leq 1.15\,{\rm GeV/n}$, $\Phi_{\bar d}<1.9\times 10^{-4} \,{\rm m}^{-2} {\rm s}^{-1} {\rm sr}^{-1} ({\rm GeV/n})^{-1}$~\cite{Fuke:2005it}, and lies well above the expected fluxes from secondary production or from dark matter annihilation/decay processes. Interestingly, the sensitivity of experiments to the cosmic antideuteron flux will increase significantly, by more than two orders of magnitude, in the near future. The Alpha Magnetic Spectrometer Experiment (AMS-02) on board of the international space station is currently searching for cosmic antideuterons in two energy windows, $0.2\leq T\leq 0.8~{\rm GeV/n}$ and $2.2\leq T\leq 4.4~{\rm GeV/n}$, with an expected flux sensitivity after five years $\Phi_{\bar d}= 1.0 \times 10^{-6} {\rm m}^{-2} {\rm s}^{-1} {\rm sr}^{-1} ({\rm GeV/n)}^{-1}$ in both energy windows~\cite{Doetinchem}. Furthermore, starting in 2014, the balloon borne General Antiparticle Spectrometer (GAPS) will undertake a series of flights at high altitude in the Antarctica searching for cosmic antideuterons. In the first phase, a long duration balloon (LDB) flight will search for antideuterons in the range of kinetic energy per nucleon $0.1\leq T\leq 0.25~{\rm GeV/n}$, with a sensitivity $\Phi_{\bar d}=1.2\times 10^{-6} {\rm m}^{-2} {\rm s}^{-1} {\rm sr}^{-1} ({\rm GeV/n)}^{-1}$, while the second, the ultra long duration balloon (ULDB) flight will search in the range $0.1\leq T\leq 0.25~{\rm GeV/n}$, with a sensitivity $\Phi_{\bar d}=3.5\times 10^{-6} {\rm m}^{-2} {\rm s}^{-1} {\rm sr}^{-1} ({\rm GeV/n})^{-1}$~\cite{Doetinchem}. The prospected sensitivities of AMS-02 and GAPS are also shown in the plot.

\begin{figure}
\begin{center}
\includegraphics[bb=0 0 220 140,width=0.48\textwidth]{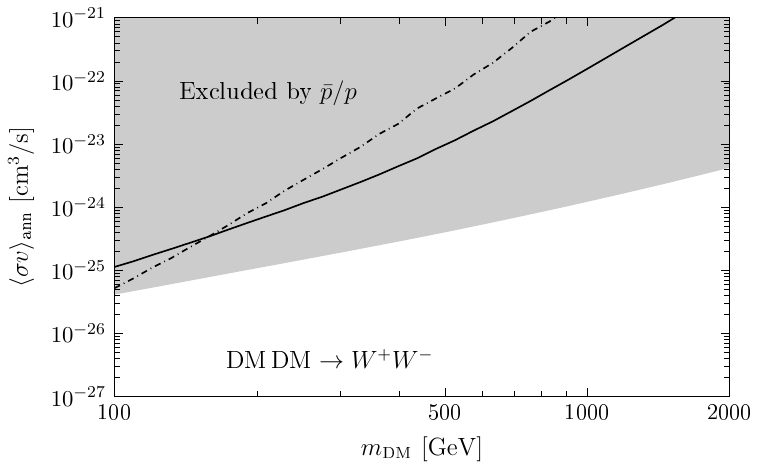}
\includegraphics[bb=0 0 220 140,width=0.48\textwidth]{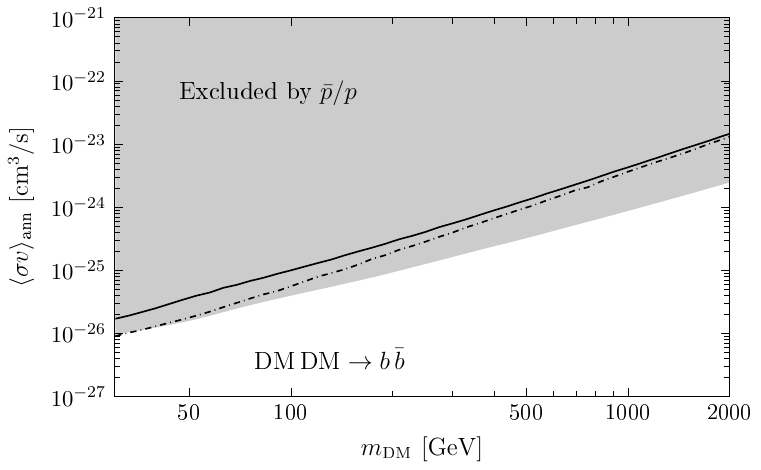}\\
\includegraphics[bb=0 0 220 140,width=0.48\textwidth]{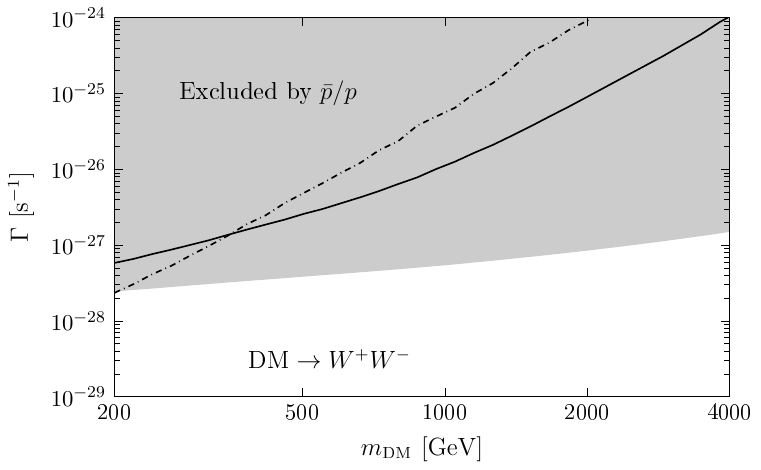}
\includegraphics[bb=0 0 220 140,width=0.48\textwidth]{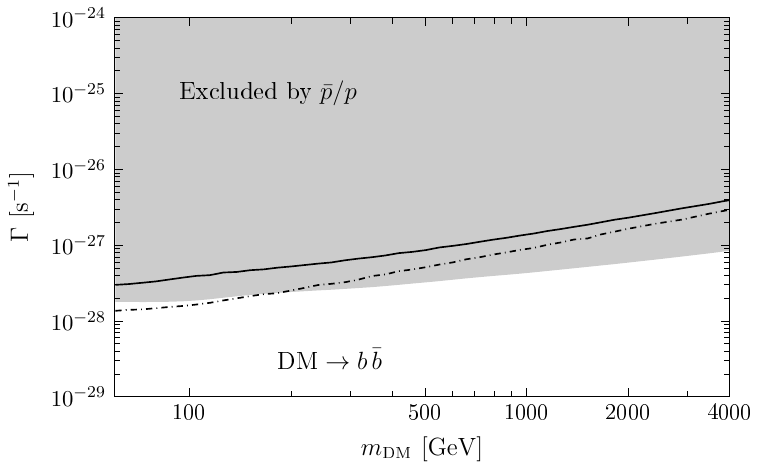}
\caption{Annihilation cross section or inverse decay rate necessary for the observation of antideuterons at the 95\% confidence level from dark matter annihilations or decays at AMS-02 (solid line) or at GAPS ULDB flight (dashed line) assuming the NFW halo profile, the MED propagation model and $p_0=192$ MeV; the shaded region is excluded from the non-observation by PAMELA of an excess in the $\bar p/p$ fraction. The top plots correspond to the annihilations ${\rm DM}\,{\rm DM}\rightarrow W^+ W^-$ (top left) and ${\rm DM}\,{\rm DM}\rightarrow b \bar b$ (top right), while the bottom plots to the decays  ${\rm DM}\rightarrow W^+ W^-$ (bottom left) and ${\rm DM}\rightarrow b \bar b$ (bottom right).}
\label{fig:x-section}
\end{center}
\end{figure}

These plots suggest that the present limits on the annihilation cross section into $W^+W^-$ or $b\bar b$ from the non-observation by PAMELA of an excess in the antiproton-to-proton fraction barely allow the observation of antideuterons at AMS-02 or GAPS. The interplay between the antiproton limits and the projected sensitivity to antideuterons at future experiments is further investigated in Fig.~\ref{fig:x-section}, where we show the dark matter annihilation cross section (top plots) and decay width  (bottom plots) necessary for the observation of antideuterons originating in dark matter annihilations or decays into $W^+W^-$ (left plots) or $b \bar b$ (right plots) at AMS-02 or GAPS ULDB with a 95\% confidence level (C.L.) excess over the background, for dark matter masses in the range 100 GeV to 2 TeV for annihilations into $W^+ W^-$ and in the range 30 GeV to 2 TeV for annihilations into $b \bar b$ (the ranges for the case of decays are twice the ranges for annihilations). In these plots we have adopted the NFW dark matter halo profile, the MED propagation model and a value of the coalescence momentum $p_0=192$ MeV. In the case of annihilation or decays into $W^+ W^-$, GAPS ULDB is the most sensitive experiment for dark matter masses below 200 GeV, while AMS-02 is more sensitive for larger masses. In this annihilation channel, the antideuteron spectrum is peaked at a kinetic energy per nucleon $T\sim m_{\rm DM}/100$, and rapidly decreases as the energy is reduced. Therefore, for large dark matter masses the number of antideuterons which fall into the AMS-02 high energy window, $2.2\leq T\leq 4.4~{\rm GeV/n}$, compensates the larger sensitivity of GAPS, which is sensitive only to lower energetic antideuterons, $0.1\leq T\leq 0.25~{\rm GeV/n}$. In contrast, the antideuteron energy spectrum for annihilation/decays into $b \bar b$ is much flatter, thus being GAPS the most sensitive experiment for the mass range of interest in this paper. We also show in the figure the 95\% C.L. excluded region of parameter space by the non-observation by PAMELA of an excess in the cosmic antiproton-to-proton fraction, adopting the background antiproton field derived in \cite{Bringmann:2006im} and the proton flux measured by AMS-01~\cite{Aguilar:2002ad}. As apparent from the plot, the PAMELA experiment already covers, for the set of assumptions listed above, the whole parameter space accessible by AMS-02 and most of the parameter space which will be explored by GAPS ULDB.

\begin{figure}
\begin{center}
\includegraphics[bb=0 0 220 140,width=0.48\textwidth]{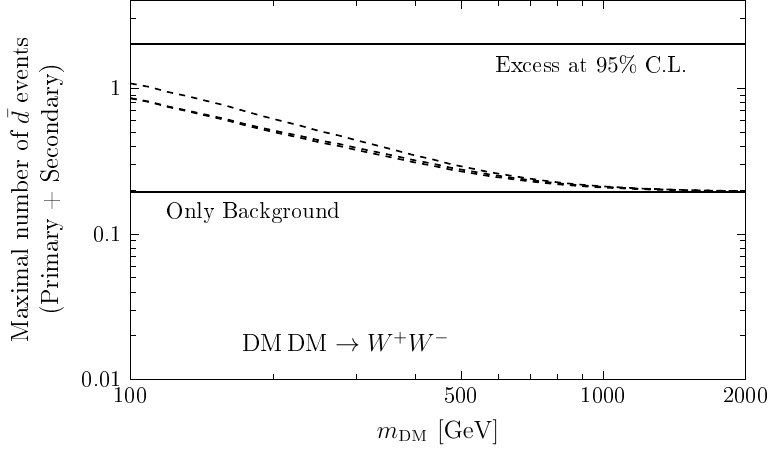}
\includegraphics[bb=0 0 220 140,width=0.48\textwidth]{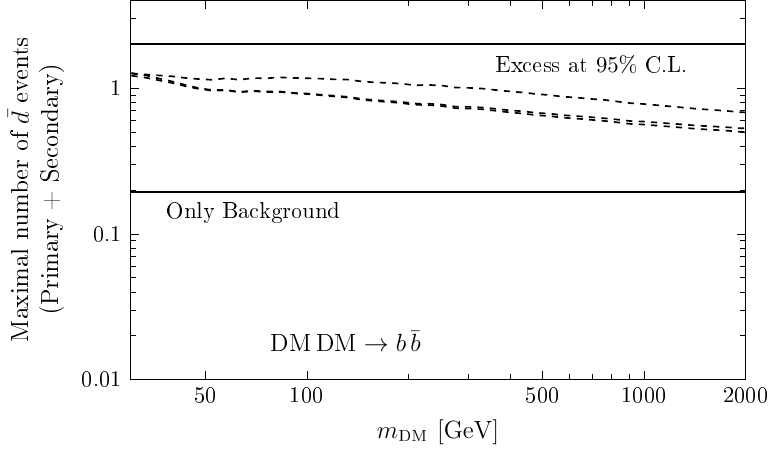}\\
\includegraphics[bb=0 0 220 140,width=0.48\textwidth]{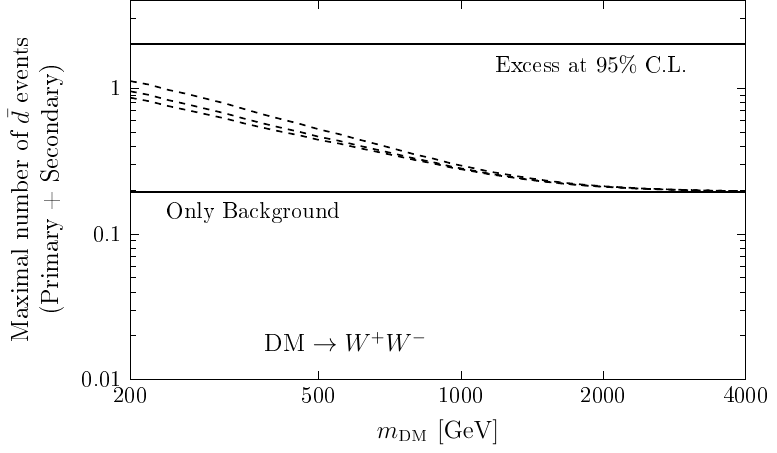}
\includegraphics[bb=0 0 220 140,width=0.48\textwidth]{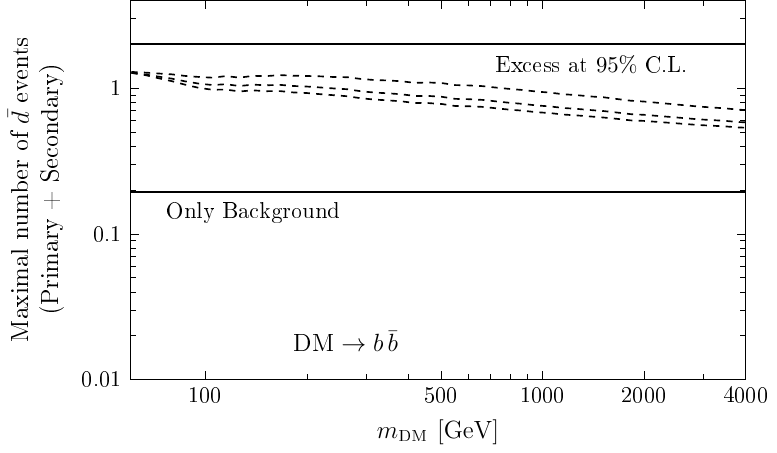}\\
\caption{Maximal number of antideuteron events expected at AMS-02 compatible with the PAMELA $\bar p/p$ data from the annihilation channels ${\rm DM}\,{\rm DM}\rightarrow W^+ W^-$ (top left) and ${\rm DM}\,{\rm DM}\rightarrow b \bar b$ (top right), and the decay channels  ${\rm DM}\rightarrow W^+ W^-$ (bottom left) and ${\rm DM}\rightarrow b \bar b$ (bottom right), assuming the NFW profile. The three dashed lines correspond (from top to bottom) to the MAX, MED and MIN propagation models. We also show the number of antideuterons expected from spallations and the total number of antideuterons corresponding to a 95\% C.L. excess.}
\label{fig:N-AMS}
\end{center}
\end{figure}

For the experimental searches it is more meaningful to calculate the upper limit on the number of events rather than the cross section. To this end, we show in Fig.~\ref{fig:N-AMS} the maximum number of antideuteron events expected at AMS-02 from the 95\% C.L. upper limit on the annihilation cross section (upper plots) or on the decay rate (lower plots) of dark matter particles into $W^+W^-$ and $b\bar b$ from the PAMELA data on the $\bar p/p$ ratio. We also show as reference the number of events expected from secondary production as well as the number of events necessary to claim an excess over the astrophysical background with a 95\% C.L. Furthermore, and in order to evaluate the impact of the choice of the propagation model in our conclusions, we show in the plot the results for the MAX, MED and MIN propagation models (from top to bottom). The weak dependence of the antideuteron flux normalized to the antiproton-to-proton fraction becomes manifest in the plot when calculating the number of events at AMS-02, and amounts to an uncertainty which is less than a 10\%. It follows from the figure that, for dark matter annihilation/decays into $W^+W^-$, the number of expected antideuteron events is below one for a wide range of dark matter masses, except when the dark matter particle is relatively light, $m_{\rm DM}\sim 200~{\rm GeV}$, where one event could be observed after 5 years of data taking. It is important to emphasize that the observation of one cosmic antideuteron at AMS-02, although a fantastic discovery in itself, cannot be interpreted as a signal of dark matter annihilation/decay since it could be an upper fluctuation of the antideuteron background. Instead, in order to claim a 95\% C.L. indication for an antideuteron excess, at least 2 events should be observed. Similar conclusions can be drawn when the dark matter particle annihilates or decays into $b\bar b$. \footnote{The number of background events just in the low energy window of AMS-02 amounts to 0.034, hence the search for dark matter particles in the window  $0.2\leq T\leq 0.8~{\rm GeV/n}$ is background free with the AMS-02 sensitivity. Nevertheless, restricting the search to this window yields a number of antideuterons from dark matter annihilations/decays which is smaller than one in the whole range of masses.}

\begin{figure}
\begin{center}
\includegraphics[bb=0 0 220 140,width=0.48\textwidth]{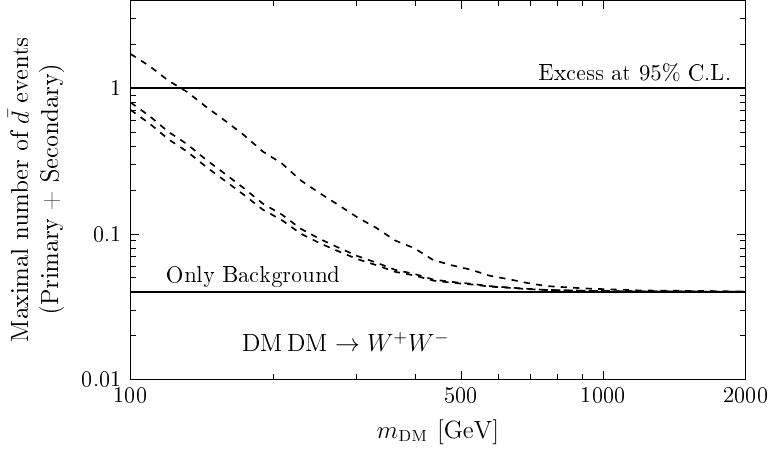}
\includegraphics[bb=0 0 220 140,width=0.48\textwidth]{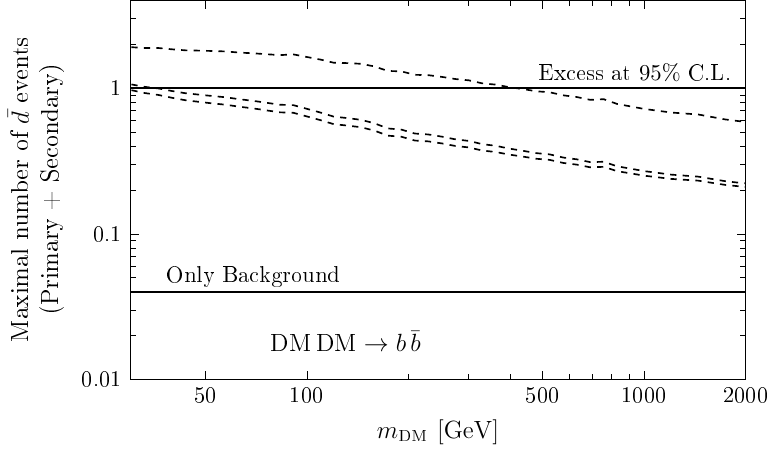}\\
\includegraphics[bb=0 0 220 140,width=0.48\textwidth]{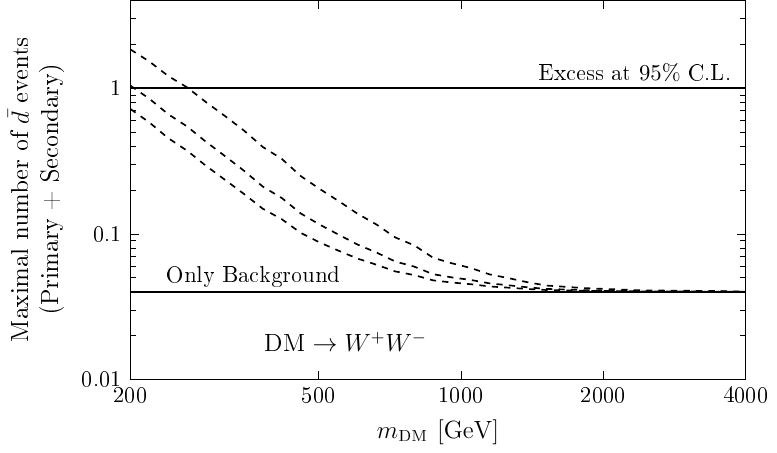}
\includegraphics[bb=0 0 220 140,width=0.48\textwidth]{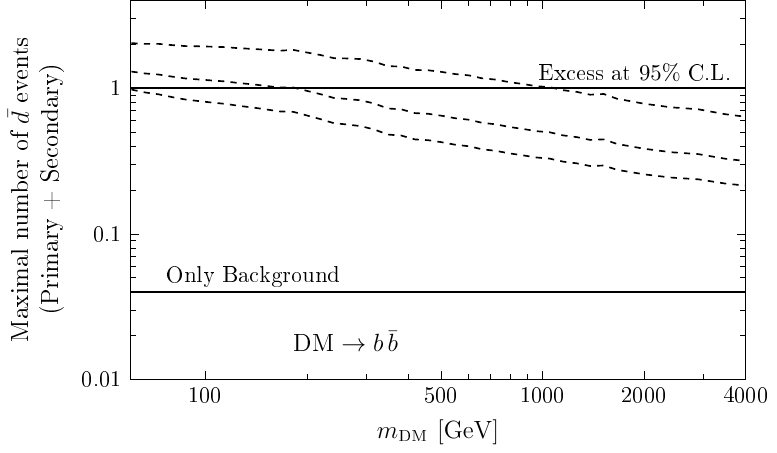}\\
\caption{The same as Fig.\ref{fig:N-AMS} but at GAPS ULDB.}
\label{fig:N-GAPS}
\end{center}
\end{figure}

The prospects for detection of antideuterons from dark matter annihilations/decays at GAPS ULDB are shown in Fig. \ref{fig:N-GAPS}. For GAPS ULDB it is expected the observation of $N_{\bar d}=0.04$ antideuterons from spallations of cosmic rays on the interstellar medium, thus being practically a background free experiment. Indeed, as shown in the figure, the observation of one single antideuteron event would suffice to assert with a 95\% C.L. the exotic origin of that antideuteron. Again, we find that for a wide range of parameters it is expected less than one antideuteron event for the MED and MIN propagation models, except for very low dark matter masses. Only for the MAX propagation model, the observation of one or two antideuteron events would be possible in view of the constraints on these annihilation/decay channels from the non-observation of an excess in the $\bar p/p$ ratio measured by PAMELA, provided $m_{\text{DM}} < 125$ GeV ($m_{\text{DM}} < 400$ GeV) for annihilations into $W^+W^-$ ($b \bar{b}$); for decaying dark matter the results are analogous, replacing $m_{\text{DM}}$ with $2 \, m_{\text{DM}}$.

It should be stressed that these conclusions apply for a value of the coalescence momentum $p_0=192$ MeV. However, as argued in section \ref{sec:coalescence}, for annihilations into $b \bar b$ the coalescence momentum could be larger for heavier dark matter particles, when the center of mass energy of the process increases. If this is the case, the antideuteron flux, which is proportional to the third power of the coalescence momentum, could be significantly enhanced, improving the prospects to detect antideuterons from dark matter annihilations/decays at AMS-02 or at GAPS. For example, if the coalescence momentum for the annihilation of dark matter particles into $b\bar b$ reaches 236 MeV, as hinted by the ZEUS data, the number of expected antideuterons would almost double the prospects presented in the right plots of Figs.~\ref{fig:N-AMS},\ref{fig:N-GAPS}.\footnote{As mentioned in footnote \ref{footnote-had}, the calculation of the background flux also suffers from hadronic uncertainties which conservatively amount to one order of magnitude. The uncertainty in the number of background events then ranges between 0.09 and 0.49 at AMS-02 and between 0.018 and 0.16 at GAPS. Hence, in the case of the largest estimate of the background, it would be necessary the observation of 3 events at AMS-02 and 2 events at GAPS to claim a 95\% C.L. indication of an antideuteron excess. In the case of the smallest estimate of the background, the number of events do not differ to the ones quoted in the main text.}

In all the plots presented in this paper we have assumed a Fisk potential $\phi_F=500$ MeV. However, data taking at future experiments will not necessarily occur during the minimum of the solar cycle, as was the case of the PAMELA experiment. However, increasing the Fisk potential in Eq.~(\ref{eq:solar-modulation}) has the effect of reducing the number of antideuterons at the top of the atmosphere, thus making our prospects more pessimistic.

\begin{figure}
\begin{center}
\includegraphics[bb=0 0 220 140,width=0.48\textwidth]{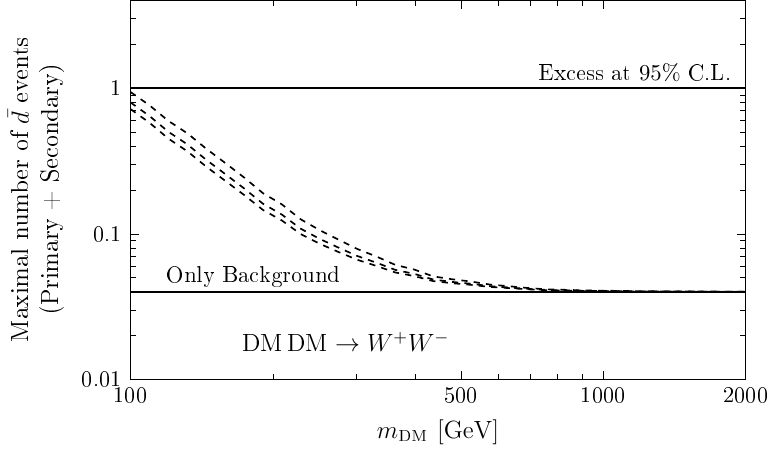}
\includegraphics[bb=0 0 220 140,width=0.48\textwidth]{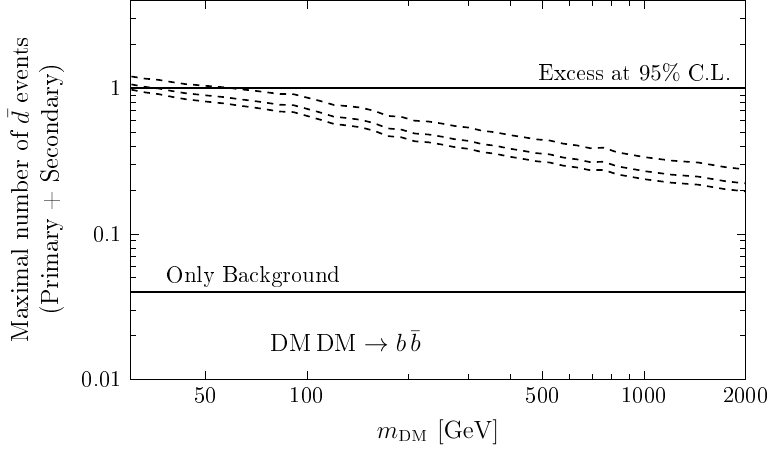}\\
\includegraphics[bb=0 0 220 140,width=0.48\textwidth]{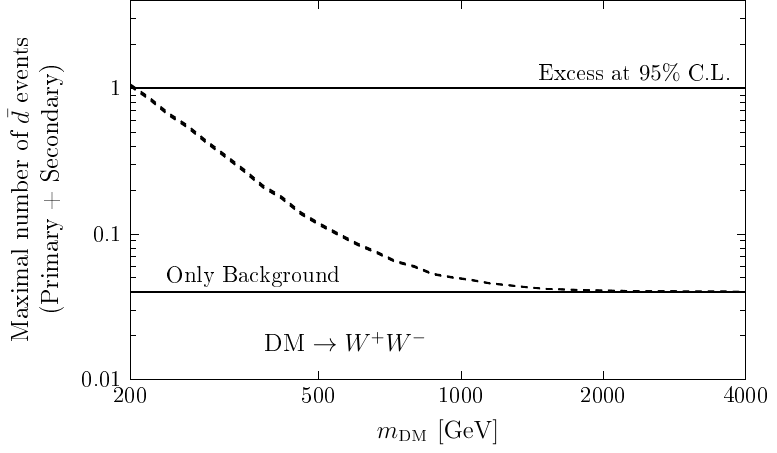}
\includegraphics[bb=0 0 220 140,width=0.48\textwidth]{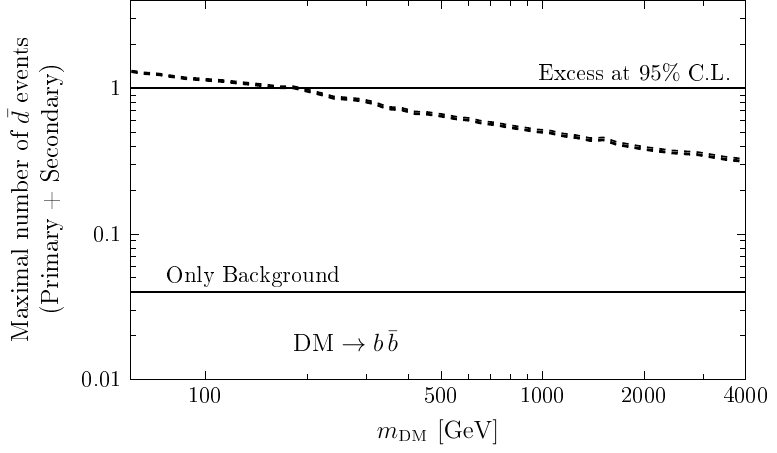}
\caption{Sensitivity to the choice of halo profile of the maximal number of antideuterons expected  from dark matter annihilations (top plots) or decays (bottom plots) into $W^+W^-$ (left plots) or $b \bar b$ (right plots) at GAPS ULDB. The three lines correspond (from top to down) to the Isothermal, the NFW and the Einasto profile. }
\label{fig:halo-dependence-GAPS}
\end{center}
\end{figure}

Lastly, we would like to briefly discuss the dependence of the results on the choice of the halo model. We show in Fig.\ref{fig:halo-dependence-GAPS} the maximum number of antideuteron events at GAPS ULDB compatible with the PAMELA antiproton-to-proton fraction for the NFW, the Einasto and the Isothermal profile; in each case we use the MED propagation model and $p_0=192$ MeV. Our results show that the uncertainty in the number of antideuterons at GAPS ULDB is smaller than 50\% in the case of annihilations and smaller than 5\% in the case of decays; the uncertainty at AMS-02 is even smaller. 

\section{Conclusions}
\label{sec:conclusions}

We have studied the prospects to observe antideuterons at AMS-02 and GAPS from dark matter annihilations or decays, in view of the stringent limits on the existence of an exotic component in the cosmic antiproton-to-proton ratio which follow from the PAMELA measurements. The antideuteron production was simulated using the coalescence model in an event-by-event Monte Carlo, being the coalescence momentum the only free parameter. To determine the value of this parameter we analyzed data from the experiments ALEPH, CLEO, CERN-ISR and ZEUS, as well as preliminary data from ALICE. We have found that the event-by-event Monte Carlo reproduces fairly well the data, however the coalescence momentum is not a universal parameter but depends on the underlying process and the center of mass energy. This result suggests that the coalescence model commonly used in the literature is oversimplified and should be extended. In this paper we have not attempted to construct an improved coalescence model but, in view of the qualitatively good description of the experimental data on antideuterons, we have used the simplest model adopting the value of the coalescence momentum inferred from the ALEPH measurements of the antideuteron multiplicity in $Z$ boson decay and we have discussed the impact of a different value of the coalescence momentum on our conclusions.

The calculation of the maximal antideuteron flux and the maximal number of antideuteron events from dark matter annihilation or decays at AMS-02 or GAPS suffers from various sources of uncertainty. To evaluate the particle physics uncertainties we have considered two representative dark matter annihilation and decay modes, namely into $W^+ W^-$ and into $b \bar b$. Furthermore, we have taken into account various astrophysical uncertainties, such as the choice of the dark matter halo model, the propagation model or the value of the Fisk potential. While the primary antiproton and antideuteron fluxes are very sensitive to the choice of the propagation model and, to a lesser extent, to the choice of the halo model, due to the fact that the antiproton and antideuteron propagation are tightly correlated, most of the uncertainties cancel out when calculating the primary antideuteron flux normalized to the primary contribution to the cosmic antiproton-to-proton ratio, giving maximum fluxes which are fairly insensitive to the astrophysical uncertainties. 

We have found that the stringent limits on the abundance of primary antiprotons in cosmic rays from the PAMELA experiment severely constrain the possibility of observing antideuterons from dark matter annihilations or decays at AMS-02 or GAPS. At AMS-02 the expected number of antideuterons from dark matter annihilation or decays is smaller than one, except when the dark matter particle is relatively light, where {\it at most} one event could be found, according to our calculations. Furthermore, at AMS-02 we expect the observation of $\sim 0.2$ secondary antideuterons, hence even if the upper limit on the number of antideuterons is saturated, the observation of one single antideuteron could not be unequivocally attributed to dark matter annihilations or decays. The conclusions for the GAPS experiment are analogous for the MIN and MED propagation model and more optimistic for the MAX propagation model.

\vspace{0.5cm}
\section*{Acknowledgements}
The authors would like to thank Miguel Pato and Philip von Doetinchem for useful discussions. This work was partially supported by the DFG cluster of excellence ``Origin and Structure of the Universe.''


\end{document}